# Assessing the Level of Autonomic Nervous Activity for Effective Biofeedback Training

*Michel KANA*

Dept. of Biomedical Informatics, Czech Technical University in Prague, Nám. Sítná 3105, 272 01 Kladno, Czech Republic

**Abstract.** *This paper proposes a prototype of a new biofeedback training based on mathematical models of cardiovascular control. For this purpose we develop a low-cost device that is able to record and process arterial pulse wave via photoplethysmograph and skin temperature on the peripheral part of the arm. A benchmark analysis of our device against a registered cardiovascular measurement system (Biopac MP35 from Biopac Inc., USA) shows that heart rate and skin temperature values delivered by our device are acceptable with a very low residual error (± 1 beat/min, ± 1°C).*

*Both measured signals are feed into a mathematical model which estimates the level of activity of both sympathetic and parasympathetic branches of the autonomic nervous system. That information is used jointly with other biological parameters for investigating the stress score of the subject. The data is processed in real-time and continuously displayed to the user for effective biofeedback training.*

*The complete solution was preliminary tested on three volunteers who used the displayed biofeedback information in order to regulate their emotional state successfully during Biofeedback training. They exhibited a significant reduction in stress score compared to three control subjects who did not used our solution during biofeedback training. This supports the benefits of biofeedback training with autonomic nervous tone assessment as effective holistic healing method.*

## Keywords

## 1. Background

Biofeedback is a learning process for gaining capability to influence physiological activities that normally happen unconsciously, such as heart function, breathing and skin temperature. Biofeedback can be effective in combination with other therapies for diverse disorders of psychical and physical function such as anxiety, psychological stress, hypertension, attention disorders, urinary incontinence or depressive disorders [1].

Biofeedback therapists use sensors for recording physiological signals such as skin temperature, muscle activity, heart rate, respiration, skin conductance, or brainwave activity. This stream of information is processed and presented in a simplified graphical or audible form that allows patients to sense and influence changes in their physiological activity in real time. The main limitation of traditional biofeedback is that subjects receive very simple audio and video feedback information [2].

We believe that assessing the level of autonomic nervous activity and feeding it back to subject's awareness will improve the effectiveness of biofeedback therapy. Autonomic control centers in the brain monitor and regulate important functions. They maintain body temperature and body osmolarity; they control reproductive function, food intake; and they influence behavior, emotions and the cardiovascular control center [3]. Assessing autonomic function is usually indirectly done by studying plasma or urinary levels of neurotransmitters, baroreflex sensitivity or heart rate variability [4]. Results are however not always reliable [5]. An attempt for improving the quantitative assessment of autonomic cardiovascular control is to combine the information obtained from the recorded physiological signal with the information derived from mathematical models [6]. Such a method is developed in this paper together with a complete hardware and software solution for effective biofeedback training.

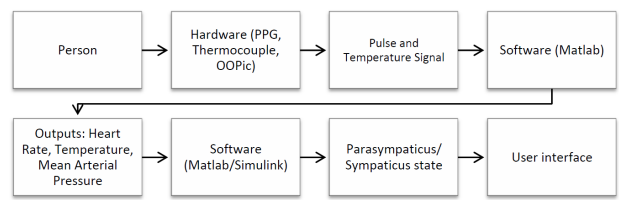

**Fig. 1.** Biofeedback solution based on mathematical models of cardiovascular neural control

## 2. Materials and Methods

As depicted in Fig. 1, our biofeedback method is based on measurement of skin temperature and pulse wave. The temperature is recorded using a thermocouple. The pulse wave is recorded using a photoplethysmograph (PPG). The data from both sensors is processed in a mathematical model which provides information about the balanced state of autonomic nervous system to the user.

### 2.1 Hardware Development

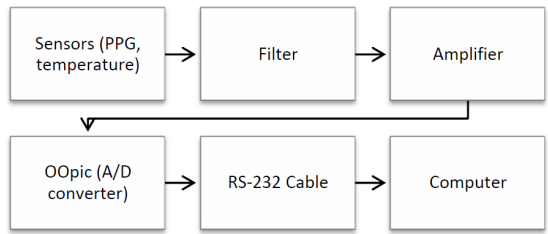

**Fig. 2.** Schematic of the hardware prototype

We use the photoplethysmographic method for heart rate measurement [7]. Light can be transmitted through tissues, which are full of small capillaries. They are filled via arterial pulsations during each heart beat. The resulting changes in volume of those vessels modify the absorption, reflection, and scattering of the light. By measuring the level of modification of the light source using a photoreceptor, it is possible to measure heart rate. As light source we used an AlGaAs (Aluminum, Gallium, Arsenide) LED, which produces a narrow-band source with a peak spectral emission at a wavelength of 850 nm. As photosensor we used a BPW42 phototransistor which is sensible to the infrared band, from 560 to 980 nm, with peak wavelength sensitivity at 830 nm. As depicted in Fig. 3, the output of a light-emitting diode is altered by tissue absorption and modulates the phototransistor. The DC level is blocked by the capacitor. A noninverting amplifier drives low impedance loads, and provides a gain of 100. Power supply of photoplethysmograph circuit is obtained from 4.5 volts batteries.

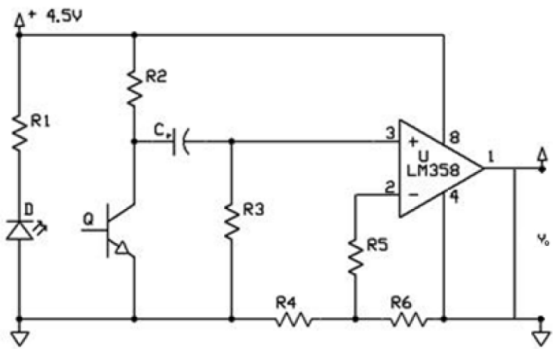

**Fig. 3.** Photoplethysmograph circuit used for heart rate measurement
(R1=100Ω, R2= 100kΩ, R3=1.6mΩ, R4=1.6mΩ, R5= 100kΩ, R6= 1kΩ, C=2μF; Diode LED IR850, 5nm, 160 nW, λ=850nm; Phototransistor BPW42, 3mm, $λ_p$=830nm)

We use the Seebeck effect for skin temperature measurement. It describes the electromotive force that exists across a junction of two dissimilar metals.

We choose the Iron – Constantan thermocouple type J with 0.25 mm diameter. It has a linear characteristic in the area around 37°C and sufficient range of values. The raw signal is 1000 times amplified in order to acquire a signal in order of volts using the instrumentation amplifier INA128 by Texas Instruments. Power supply is obtained from ±9 volts batteries.

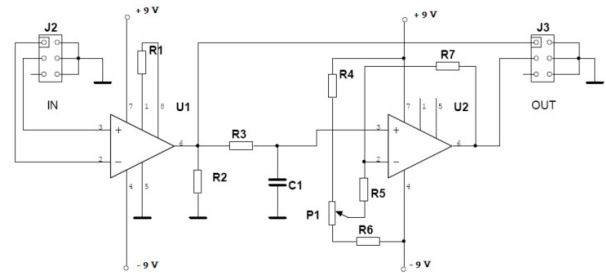

**Fig. 4.** Thermocouple circuit used for temperature measurement
(R1=56Ω, R2=10kΩ, R3=1kΩ, R4=500Ω, R5=10kΩ, R6=500Ω, R7=10kΩ, P1=10kΩ; C1=47nF, C2=10pF; U1 INA128, U2 TL074)

The Object Oriented Programmable Integrated Circuit (OOPic by Savage Innovations) microcontroller was used for the purpose of 5 volts A/D conversion and RS-232 serial communication with the computer.

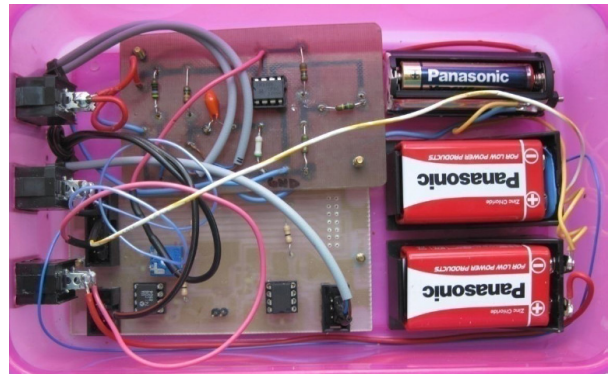

**Fig. 5.** Box with boards, power supply and supply switches

The final hardware prototype is shown in Fig. 5. We validated our hardware solution against the well-known Biopac MP35 system, developed by Biopac Systems Inc., which is a registered system for physiological data acquisition and analysis. Simultaneous data streams from both Biopac and our systems were statistically analyzed and compared. The mean heart rate obtained by Biopac was almost the same, with maximum one beat per minute error, as the mean heart rate obtained by our system. The skin temperature delivered by Biopac corresponded to the skin temperature delivered by our system.

### 2.2 Software Development

A schematic representation of our software prototype is shown in Fig. 6. When a biofeedback training session is started by a user, the *COM Reader* opens the COM port. Raw data stream comes from the OOPic microcontroller in the format: *Data Separator – Pulse Value – Temperature Value*. As a separator we used the number '2' that is not engaged in both signals. COM Reader recognizes the Data

Separator and sends the next two bytes to *Buffer Manager* that stores pulse and temperature values. Buffer Manager also immediately sends the data to *Plot Manager* for display in real-time. After a time window of 10 seconds is elapsed, the content of the buffer is sent to *Pulse Processor* for processing. The Pulse Processor detects peaks in the arterial pulse waveform and calculates the heart rate (HR). Mean arterial pressure (MAP) is estimated from the pulse amplitude using the equation (1) which was proposed in [8]. The temperature values (T) are converted from volts values to °C according the table for thermocouple type J. HR, MAP and T are sent to Plot Manager for display.

$$MAP = a \cdot e^{b \cdot A} + c \cdot e^{d \cdot A} \quad (1)$$

where the unknown parameters a, b, c and d can be determined by fitting correct blood pressure values to pulse amplitudes during a calibration phase.

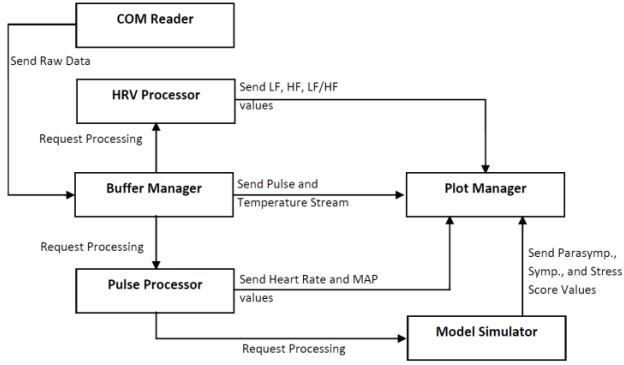

**Fig. 6.** Structure of a software solution for biofeedback training

After a time window of 30 seconds is elapsed, the HR values corresponding to the buffer content are sent to the *HRV Processor* that performs a heart rate variability analysis (HRV) in the frequency band [0.015 – 0.4] Hz by calculating the power of LF, HF components and their ratio LF/HF. The obtained HRV parameters are normalized over the total power, and then sent to Plot Manager for display. The HR, MAP, T values corresponding to the buffer content are also sent to the *Model Simulator*, that simulates a mathematical model of cardiovascular neural control (see section 2.3 of this paper) and estimates the level of activity of sympathetic and parasympathetic branches as well as the so-called stress score. All three values are sent to Plot Manager for display.

A graphical user interface was developed in the Matlab environment according to user requirements gathered during on-site research with 30 respondents from a private relaxation center and a university (see Fig. 7). It displays measured and processed physiological data as feedback for the user. This included a percentage of stress score, with an appropriate color (red for high and green for low stress score) and a smiley icon (smiling for low, satisfied for normal and angry for high stress score).

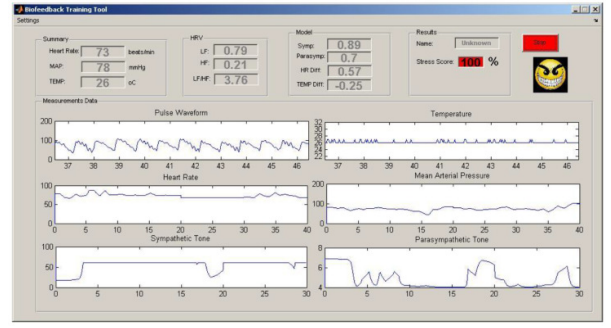

**Fig. 7.** User interface for biofeedback training with autonomic nervous activity assessment

### 2.3 Mathematical Model for Biofeedback

The level of activity of the autonomic nervous system can be estimated using mathematical models of cardiovascular control. The main component of such a model would be the baroreflex or baroreceptor loop, which stabilizes blood pressure by the mean of neural negative feedback with three components: an afferent or sensory component, an efferent component corresponding to autonomic nervous control and an effector component for modulating heart rate and vascular resistance [9]. We selected model equations from Ursino [10] because of their simplicity and their acceptability in the scientific community. We adapted the equations for our biofeedback purpose as presented below.

The *Sensory Component* consists of mechanoreceptor cells located in the carotid sinus and aorta arch walls responding to pressure by stretching. Their activity is modeled using a linear derivative first-order dynamic block and a sigmoidal static block as follows.

$$f_{ab} = \frac{f_{ab,min} + f_{ab,max} \cdot e^{\frac{\tilde{P}-P_n}{k_{ab}}}}{1 + e^{\frac{\tilde{P}-P_n}{k_{ab}}}} \quad (2)$$

$$\tau_{pb} \cdot \frac{d\tilde{P}}{dt} = P_{as} + \tau_{zb} \cdot \frac{dP_{as}}{dt} - \tilde{P}$$

where $f_{ab}$ is the firing rate of the carotid baroreceptors in response to the pressure sensed $\tilde{P}$ which depends on the arterial pressure $P_{as}$; $f_{ab,max}, f_{ab,min}$ are the upper and lower saturation levels of firing rate with values 2.52 Hz and 47.78 Hz; $\tau_{pb}, \tau_{zb}$ are the time constants for the real pole and the real zero in the linear dynamic block with values 2.07 sec and 6.37 sec; $P_n$ is a model parameters, related to the value of baroreceptor pressure at the central point of the sigmoidal function; $k_{ab}$ is a model parameter, with the dimension of pressure, related to the slope of the static function at the central point.

The *Efferent Component* consists of sympathetic and vagal nerves which project to blood vessels and heart. Their activity is modeled as exponential trends using weighted sum of afferent inputs from baroreceptors as follows.

$$f_s = \begin{cases} f_{es,\infty} + (f_{es,0} - f_{es,\infty}) \cdot e^{k_{es} \cdot (-W_{b,s} \cdot f_{ab} - \theta_s)} & if\ f_s < f_{es,max} \\ f_{es,max} & if\ f_s \geq f_{es,max} \end{cases}$$

$$f_v = \frac{f_{ev,0} + f_{ev,\infty} \cdot e^{\frac{f_{ab} - f_{ab,0}}{k_{ev}}}}{1 + e^{\frac{f_{ab} - f_{ab,0}}{k_{ev}}}} - \theta_v \quad (3)$$

$$\bar{f}_s = \frac{f_s}{f_{es,max}}$$

$$\bar{f}_v = \frac{f_v}{f_{ev,\infty}}$$

where $f_s$ is the firing rate of efferent sympathetic fibers to vessels and heart; $f_v$ is the firing rate of efferent vagal fibers to heart; $k_{es}, f_{es,max}, f_{es,\infty}, k_{ev}, f_{ev,\infty}$ are constants with values 0.06, 60 Hz, 2.1 Hz, 7 Hz, 2.1 Hz and 3.15 Hz; $W_{b,sp}$ is a synaptic weight with value 0.3; $\theta_s, \theta_v$ are offset terms for sympathetic and vagal neural activation with values -49 Hz and -0.68 Hz. $f_{es,0}, f_{ev,0}$ are model parameters, related to the intrinsic firing rate of sympathetic and vagal nerves respectively at the absence of any sensory input. $\bar{f}_s$ is the so-called **sympathetic tone**, i.e. a quantification of sympathetic activity in the normalized range [0 .. 1]. Similarly, $\bar{f}_v$ is the **parasympathetic tone**.

The *Effectors Component* consists of the heart sinoatrial node and peripheral blood vessels. The changes in skin temperature as response to sympathetic activity are modeled to include a pure latency, a monotonic logarithmic static function, and low-pass first order dynamics.

$$T_s(t) = \Delta T_s(t) + T_{s,0}$$
$$\frac{d\Delta T_s}{dt} = \frac{1}{\tau_{T_s}} \cdot (-\Delta T_s + \sigma_{T_s}) \quad (4)$$
$$\sigma_{T_s} = \begin{cases} G_{T_s} \cdot ln[f_s(t - D_{T_s}) - f_{es,min} + 1] & if\ f_s \geq f_{es,min} \\ 0 & if\ f_s < f_{es,min} \end{cases}$$

where $T_s$ is the skin temperature in ºC, related to the resistance of peripheral vessels with $T_{s,0}$ being its baseline value as model parameter; $\tau_{T_s}$ is a time constant with value 6; $G_{T_s}$ is a constant gain factor with value 1.94; $D_{T_s}$ is the time delay for sympathetic response to take effect with value 5 sec; $f_{es,min}$ is a threshold for sympathetic stimulation with value 2.66 Hz.

The changes in the heart rate depend of both sympathetic and vagal drive as follows.

$$HR = \frac{60}{\Delta HR_S + \Delta HR_V + HR_0}$$
$$\frac{d\Delta HR_S}{dt} = \frac{1}{\tau_{HR_S}} \cdot (-\Delta HR_S + \sigma_{HR_S})$$
$$\frac{d\Delta HR_V}{dt} = \frac{1}{\tau_{HR_V}} \cdot (-\Delta HR_V + \sigma_{HR_V}) \quad (5)$$
$$\sigma_{HR_S} = \begin{cases} G_{HR_S} \cdot ln[f_s(t - D_{HR_S}) - f_{es,min} + 1] & if\ f_s \geq f_{es,min} \\ 0 & if\ f_s < f_{es,min} \end{cases}$$
$$\sigma_{HR_V} = G_{HR_V} \cdot f_v(t - D_{HR_V})$$
$$HR_0 = \frac{1}{1.97 - 9.5 \cdot 10^{-3} \cdot age}$$

where $HR$ is the heart rate in beats/min with $HR_0$ being the intrinsic duration of 1 cardiac cycle, depending on the $age$; $\tau_{HR_S}, \tau_{HR_V}$ are a time constants with values 2 and 1.5; $G_{HR_S}, G_{HR_V}$ are constant gain factors with values -0.13 and 0.09; $D_{HR_S}, D_{HR_V}$ are the time delays for sympathetic and parasympathetic responses to take effect with values 2 sec and 0.2 sec.

## 2.4 Experiments

Three healthy men and three healthy women (age 37±13 years) with different employments and stress environments were divided into two randomly selected groups. The first one undergoes whole biofeedback training and the second one was a non-training control group. The control and trained group were managed according to training plans described in the Fig. 8 and 9. All subjects were instrumented with the hardware device we have developed for arterial pulse and skin temperature measurement (see section 2.1).

The goal of the training for both groups was to improve the ability to calm down and relax. The methods for relaxation were deep diaphragmatic breathing and enhanced peripheral capillary return. Deep diaphragmatic breathing was performed at a rate of 6 cycles per minute in a comfortable sitting position with one hand placed on the upper chest and the other just below your rib cage in order to feel the diaphragm moving. The subjects were instructed to tighten the stomach muscles, letting them fall inward as they exhale through pursed lips, while the hand on the upper chest remains as still as possible. Enhanced peripheral capillary return was upheld by repeating autogenic phrases such as "I have warm stones in my hands which heat them". Relaxation was further enhanced in the training group by gentle massage and few minutes resting in yoga asana Savasana which has the same sedative effect as some kind of antidepressants. The control group did not undergo massage and yoga.

The principle of training was based on scoring the overall stress state in order to provide the learner with combined information about his or her heart rate, skin temperature, heart rate variability, level of sympathetic or parasympathetic activity and stress score in real time. The control group did not have access to that combined information. Instead control subjects had to achieve the biofeedback goal without being presented with any information about their physiological state.

Each biofeedback session had 4 phases. The pure measurement phase is for the subject to get familiarized with the experimental setup. The deep diaphragmatic breathing phase is preparation phase where subjects get relaxed by activating their parasympathetic system via deep breathing and fictive image of warm hands. The resting phase is a pause prior to the final phase. The Self-calming phase is the active biofeedback session, where the subject consciously tries to take control of its physiological variables while reducing his stress score, sympathetic tone and low frequency components in heart rate variability, and increasing his parasympathetic tone and high frequency components in heart rate variability at the same time.

| Duration | Activity | | | |
|---|---|---|---|---|
| 20 minutes | Introduction, explanation of biofeedback, measured outcomes, diaphragmatic breathing and breathing practice | | | |
| 7 minutes | Biofeedback training with our biofeedback device; emphasis on breathing, image of warm hands | | | |
| | | Phase | Duration | Activity |
| | | Phase 1 | 1. – 90. sec. | Pure measurement |
| | | Phase 2 | 91. – 240. sec. | Deep diaphragmatic breathing, image of warm hands |
| | | Phase 3 | 241. – 300. sec. | Resting |
| | | Phase 4 | 301. – 400. sec. | Self-calming |
| 10 minutes | Massage of nape and neck; emphasis on body ease, deep breath, clear mind (without biofeedback device) | | | |
| 5 minutes | Lying in Savasana | | | |
| 7 minutes | Biofeedback training with our biofeedback device; emphasis on breathing, image of warm hands | | | |
| | | Phase | Duration | Activity |
| | | Phase 1 | 1. – 90. sec. | Pure measurement |
| | | Phase 2 | 91. – 240. sec. | Deep diaphragmatic breathing, image of warm hands |
| | | Phase 3 | 241. – 300. sec. | Resting |
| | | Phase 4 | 301. – 400. sec. | Self-calming |
| 5 minutes | Time allowance | | | |

**Fig. 8.** Experimental schedule of the training group.

| Duration | Activity | | | |
|---|---|---|---|---|
| 20 minutes | Introduction, explanation of biofeedback, measured outcomes | | | |
| 7 minutes | Biofeedback training (measurement) without our biofeedback device; emphasis on breathing, image of warm hands | | | |
| | | Phase | Duration | Activity |
| | | Phase 1 | 1. – 90. sec. | Pure measurement |
| | | Phase 2 | 91. – 240. sec. | Deep breathing, image of warm hands |
| | | Phase 3 | 241. – 300. sec. | Resting |
| | | Phase 4 | 301. – 400. sec. | Self-calming |
| 5 minutes | Time allowance | | | |

**Fig. 9.** Experimental schedule of the control group.

## 2.5 Parameters estimation

Beat-to-beat heart rate ($HR_{meas}$) and skin temperature ($T_{meas}$) were directly extracted from the pulse plethysmograph signal which was recorded during the all four phases of the biofeedback training. The obtained HR time series was analyzed in the frequency domain using Fourier transformation in order to calculate the normalized power of low and high frequency components of heart rate variability (LF, HF). Mean arterial pressure ($MAP_{meas}$) was estimated from the amplitude of peaks in the pulse plethysmograph signal using the mathematical equation (1).

In order to compute the sympathetic tone $\overline{f}_s$ and the parasympathetic tone $\overline{f}_v$, the mathematical model described in section 2.3 was implemented in the Simulink software environment and model parameters were estimated by formulating and resolving a least-squares optimization problem minimizing the errors between measured heart rate, and skin temperature $HR_{meas}, MAP_{meas}, T_{meas}$ and simulated heart rate, and skin temperature $HR, P_{as}, T_s$. The corresponding cost function, i.e. the discrepancy between measured and simulated values is defined as optimization criteria as follows, where $n$ is the number of data points.

$$J = \sum_{i=1}^{n} \left[HR_{meas}(i) - HR(i)\right]^2 + \left[MAP_{meas}(i) - P_{as}(i)\right]^2 + \left[T_{meas}(i) - T_s(i)\right]^2 \quad (5)$$

The cost function was minimized in the Matlab software environment with parameters values shown in table 1 for a representative subject.

| Model Parameter | Value |
|---|---|
| $P_n$ | 92 mmHg |
| $k_{ab}$ | 11.7 mmHg |
| $f_{es,0}$ | 16.1 Hz |
| $f_{ev,0}$ | 3.14 Hz |
| $T_{s,0}$ | 25 ºC |
| $G_{HR_s}$ | -0.13 |
| $G_{HR_v}$ | 0.09 |

**Tab. 1.** Typical values of model parameters

The simulated sympathetic and parasympathetic tones were combined with variables extracted from experimental data in order to calculate the stress score using the formula (6) below. Sympathetic tone, high fluctuations in heart rate, average skin temperature close to 37 ºC and average heart rate less than 70% of the intrinsic heart rate contribute to lower stress score, meaning that the subject is fully relaxed under predominant activity of the parasympathetic branch of the autonomic nervous system, compare to minimal activity of the sympathetic branch.

$$SC = 100 \cdot \left[ 1 - \frac{\bar{f}_v + HF - \bar{f}_s - LF + \frac{HR_0 - \frac{60}{HR_{meas}}}{0.7 \cdot HR_0} + \frac{T_{s,0}}{37}}{3} \right] \quad (6)$$

## 3. Results

In order to evaluate the effectiveness of our biofeedback method which is based on autonomic nervous assessment, we compared the change in different variables between two consecutive phases of the biofeedback training for the control group, the trained group before relaxation and the trained group after relaxation. Especially we are interested in the capability of the trained group to more effectively reduce its stress score and increase its parasympathetic tone during self calming (phase 4).

Heart rate decreased during deep diaphragmatic breathing (phase 2) for all groups. This trend was also observed in the trained group during self calming. The control group displayed an increasing heart rate instead, as shown in Fig. 10.

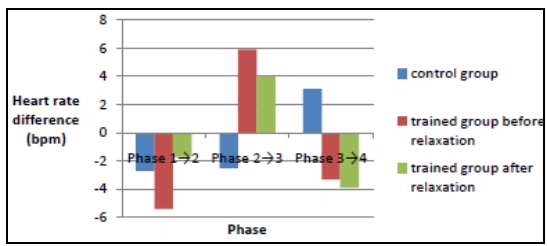

**Fig. 10.** Change in the average heart rate.

The changes in the normalized power of low frequency components of heart rate variability, which reflect sympathetic activity, are the most expressive for the trained group. Subjects in the trained group had visual information about their autonomic nervous activity and could reduce sympathetic activity in phase 4 (see Fig. 11) while enhancing vagal activity. In Fig. 12, the power of high frequency components of heart rate variability, which are known to reflect vagal activity and a state of relaxation, significantly increases. The control group did not exhibit any change.

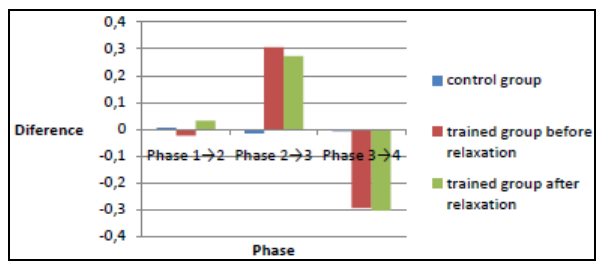

**Fig. 11.** Changes in the normalized power of low frequency components of heart rate variability.

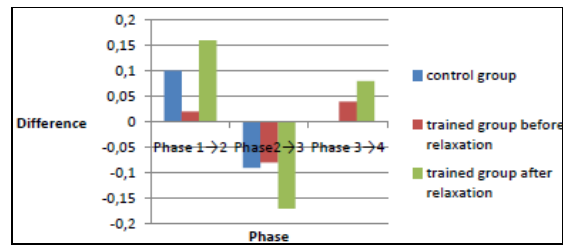

**Fig. 12.** Changes in the normalized power of high frequency components (HF) of heart rate variability.

As shown in Fig. 13, the control group had the highest stress score. Stress decreased during biofeedback training with our software solution. Relaxation prior to biofeedback training provided additional reduction of stress score.

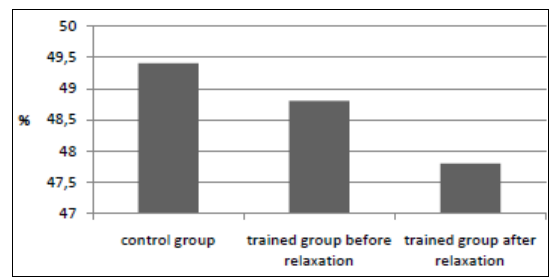

**Fig. 13.** Stress score achieved during self calming.

Differences between the control and trained group are obvious as explained in previous paragraphs. The trained groups show better results (reduced stress score, reduced sympathetic tone, increased parasympathetic activity). It was also observed that passive relaxation techniques are valuable for effective biofeedback training. We believe that our complete and functional solution for biofeedback training can be used in alternative medicine, i.e. therapy with limited adverse effects.

Future work could be in the design of a better photosensor gripping on the finger in order to reduce noise in the pulse signal. Moreover we should improve the mathematical model of autonomic nervous control and include additional neural pathways that are involved in sympathetic and parasympathetic function. We also plan to test the solution on more volunteers in order to bring a solid statistical proof of its significant effectiveness in biofeedback training. A further room of improvement would be the inclusion of environmental parameters to our solution such as ambient temperature, humidity and atmospheric pressure which can be very influential on the autonomic nervous system of human subjects.

## References

[1] Schwartz, M and Andrasik, F., *Biofeedback: A Practitioner's Guide.* 3rd Edition. s.l. : The Guilford Press, 2005. ISBN 1593852339.

[2] Arns, M, et al., "Efficacy of neurofeedback treatment in ADHD: the effects on inattention, impulsivity and hyperactivity: a meta-analysis." Clin EEG Neurosci., 2009, Issue 3, Vol. 40, pp. 180-9.

[3] Unglaub, D., *Human Physiology: An Integrated Approach.* [ed.] 4th Edition. s.l. : Pearson, 2009. 0805368493.

[4] Malik, M., "Heart Rate Variability. Standards of Measurement, Physiological Interpretation and Clinical Use." Circulation, 1996, Issue 5, Vol. 93, pp. 1043-1065.

[5] Goldberger, J J, et al., "Dissociation of heart rate variability from parasympathetic tone." Am J Physiol Heart Circ Physiol, 1994, Issue 5, Vol. 266, pp. H2152-H2157.

[6] Batzel, J, et al., *Cardiovascular and Respiratory Systems – Modeling, Analysis and Control.* Philadelphia : Society for Industrial and Applied Mathematics, 2007. pp. 6-16. ISBN 0-89871-617-9.

[7] Allen, J., "Photoplethysmography and its application in clinical physiological measurement." Physiol Meas, 2007, Issue 3, Vol. 28, pp. R1-39.

[8] Shaltis, P, Reisner, A and Asada, H., "Calibration of the photoplethysmogram to arterial blood pressure: capabilities and limitations for continuous pressure monitoring." 2005. Conf Proc IEEE Eng Med Biol Soc. Vol. 4, pp. 3970-3.

[9] Ottesen, J T., "Modelling of the baroreflex-feedback mechanism with time-delay." Journal of Mathematical Biology, 1997, Issue 1, Vol. 36, pp. 41-63.

[10] Ursino, M., "Interaction between carotid baxoregulation and the pulsating heart: a mathematical model." Am J Physiol Heart Circ Physiol, 1998, Vol. 275, pp. 1733-1747.